%% file: Corr-SmartMTD.tex
\documentclass[sigconf]{acmart}

\usepackage{booktabs} 

\usepackage{graphics}
\usepackage{subfig}
\usepackage{comment}
\usepackage{multirow}
\usepackage{xcolor}
\usepackage[linesnumbered,ruled,vlined]{algorithm2e}

\SetCommentSty{mycommfont}
\usepackage{url}

\begin{document}
\title{SmartMTD: A Graph-Based Approach for Effective \\ Multi-Truth Discovery}

\author{Xiu Susie Fang}
\affiliation{%
  \institution{Department of Computing, Macquarie University}
  \city{Sydney} 
  \state{NSW} 
  \postcode{2109}
  \country{Australia}
}
\email{xiu.fang@students.mq.edu.au}

\author{Quan Z. Sheng}
\affiliation{%
  \institution{Department of Computing, Macquarie University}
  \city{Sydney} 
  \state{NSW} 
  \postcode{2109}
   \country{Australia}
}
\email{michael.sheng@mq.edu.au}

\author{Xianzhi Wang}
\affiliation{%
  \institution{School of Computer Science and Engineering, The University of New South Wales}
  \city{Sydney} 
  \state{NSW} 
  \postcode{2052}
    \country{Australia}
  }
\email{xianzhi.wang@unsw.edu.au}

\author{Anne H.H. Ngu}
\affiliation{%
  \institution{Department of Computer Science, Texas State University}
  \city{San Marcos} 
  \state{TX} 
  \postcode{78666}
    \country{USA}
  }
\email{angu@txstate.edu}

\renewcommand{\shortauthors}{X. S. Fang et al.}

\begin{abstract}
The Big Data era features a huge amount of data that are contributed by numerous sources and 
used by 
many critical data-driven applications. Due to the varying reliability of sources, it is common to see conflicts among the multi-source data, making it difficult to determine which data sources to trust. Recently, \emph{truth discovery} has emerged as a means of addressing this challenging issue by determining data veracity jointly with estimating the reliability of data sources. A fundamental issue with current truth discovery methods is that they generally assume only one true value for each object, while in reality, objects may have multiple true values. In this paper, we propose a graph-based approach, called \emph{SmartMTD}, to unravel the truth discovery problem beyond the single-truth assumption, or the multi-truth discovery problem. SmartMTD models and quantifies two types of source relations to estimate source reliability precisely and to detect malicious agreement among sources for effective multi-truth discovery. In particular, two graphs are constructed based on the modeled source relations. They are further used to derive the two aspects of source reliability (i.e., \emph{positive precision} and \emph{negative precision}) via random walk computation. Empirical studies on two large real-world datasets demonstrate the effectiveness of our approach. 
\end{abstract}

%
%




\maketitle

\input{SmartMTD}

\bibliographystyle{ACM-Reference-Format}
\bibliography{reference} 

\end{document}

%% file: SmartMTD.tex
\section{Introduction}
Nowadays, data are created at an unprecedented rate through various channels over the Web, such as blogs, social networks, discussion forums, and crowd-sourcing platforms. While the \emph{Big Data} holds the potential to revolutionize many aspects of the modern society, it is often observed that multiple sources provide conflicting descriptions on the same objects, due to typos, out-of-date data, missing records, or erroneous entries~\cite{gao2015truth,dong2009integrating,xiu2017value,wang2016empowering}.
Such conflicts may cause considerable damage and financial loss in many applications, such as healthcare systems when the data are used for drug recommendation or stock markets when that data are used for stock price prediction~\cite{benslimane2015uncertain}. Given large-scale data, since it is unrealistic to determine manually which records of data are true, \emph{truth discovery} has emerged as a fundamental technique of estimating data veracity by resolving the conflicts in multi-source data automatically.

Until now, considerable research efforts have been conducted to solve the truth discovery problem. Most of them compute source reliability and value veracity alternatively and iteratively from each other, and the existing methods~\cite{pasternack2010knowing,yin2008truth,galland2010corroborating,li2014resolving,dong2009integrating} consider various factors such as data types, source dependency, source quality to facilitate truth discovery, they commonly assume that each object has exactly one true value (i.e., {\em the single-truth} assumption). However, in the real world, multi-valued objects widely exist, such as the children of a person or the authors of a book. 
Although the previous methods can deal with multi-valued objects by simply regarding a set of values provided by a source on a single object as a joint single value, whereby the truth can be identified as the most confident value set among all the value sets provided by all sources, the value sets provided by different sources are generally correlated: there may be overlaps between the value sets claimed by two sources on the same object, indicating that the two sources may not totally vote against each other on the object. For example, a source may claim ``Daniel Radcliffe, Emma Watson, Rupert Grint'' while another source may claim ``Daniel Radcliffe, Emma Watson'' as the cast of the movie ``Harry Potter''. Apparently, the latter set is covered by the former and therefore partially supports the former set. Neglecting this implication could significantly degrade the accuracy of truth discovery. 

Another drawback of the previous single-truth discovery methods is that they usually measure source quality by a single parameter (e.g., precision or accuracy) while overlooking the important distinction between two aspects of quality, namely, false negatives and false positives. This distinction, however, is important for multi-valued objects, as some sources may tend to provide erroneous values, making more false positives, while some other sources may incline to provide partial true values without erroneous values, making more false negatives. By lump together the two types of errors in a single measure, the previous single-truth discovery methods cannot distinguish between those two types of sources. Conversely, it is crucial to 
identify the complete true values for multi-valued objects to consider these two different types of errors comprehensively in measuring source reliability. After realizing those features, several approaches~\cite{zhao2012bayesian,wang2015integrated,wang2016implications,wang2016empowering,wan2016uncertainty}
 have been proposed to tackle multi-valued objects, but they ignore the object distributions and complex source relations in their data model, rendering the problem of truth discovery for multi-valued objects, a.k.a., the multi-truth discovery (MTD) problem, still far from being solved.

In this paper, we focus on the MTD problem. In a nutshell, we make the following main contributions:
\begin{itemize}
\item We propose a graph-based model, called \emph{SmartMTD}, as an overall solution to the MTD problem. This model incorporates two important implications, namely \emph{source relations} and \emph{object popularity}, for better truth discovery.

\item We propose to model two-sided relations among sources, 
and graphs are then constructed to capture source features. Specifically, we use $\pm$\emph{supportive agreement graph}s to capture source authority features and two-sided source precision, and $\pm$\emph{malicious agreement graph}s to quantify source dependence degrees. Random-walk computation is applied to both types of graphs to estimate source reliability and dependence degrees. We further distinguish source reliability by differentiating objects by their popularity, to minimize the number of audiences misguided by false values.

\item We conduct extensive experiments to demonstrate the effectiveness of our proposed approach via comparison with the state-of-the-art baseline methods on two real-world datasets.
\end{itemize}

The rest of the paper is organized as follows. We discuss the observations that motivate our work and formulate the multi-truth discovery problem in Section~\ref{sec:Preliminaries}. Section~\ref{sec:Approach} presents our approach and the incorporated implications. We report our experiments and results in Section~\ref{sec:Experiments}. Section~\ref{sec:Related_Work} reviews the related work, and Section~\ref{sec:Conclusion} provides some concluding remarks.

\section{Preliminaries}
\label{sec:Preliminaries}
\subsection{Observations}
\label{subsec:Observations}
\begin{figure}[!tb]
\centering
\subfloat[][\scriptsize{Book-Author Dataset}]{\includegraphics[width=1.7in]{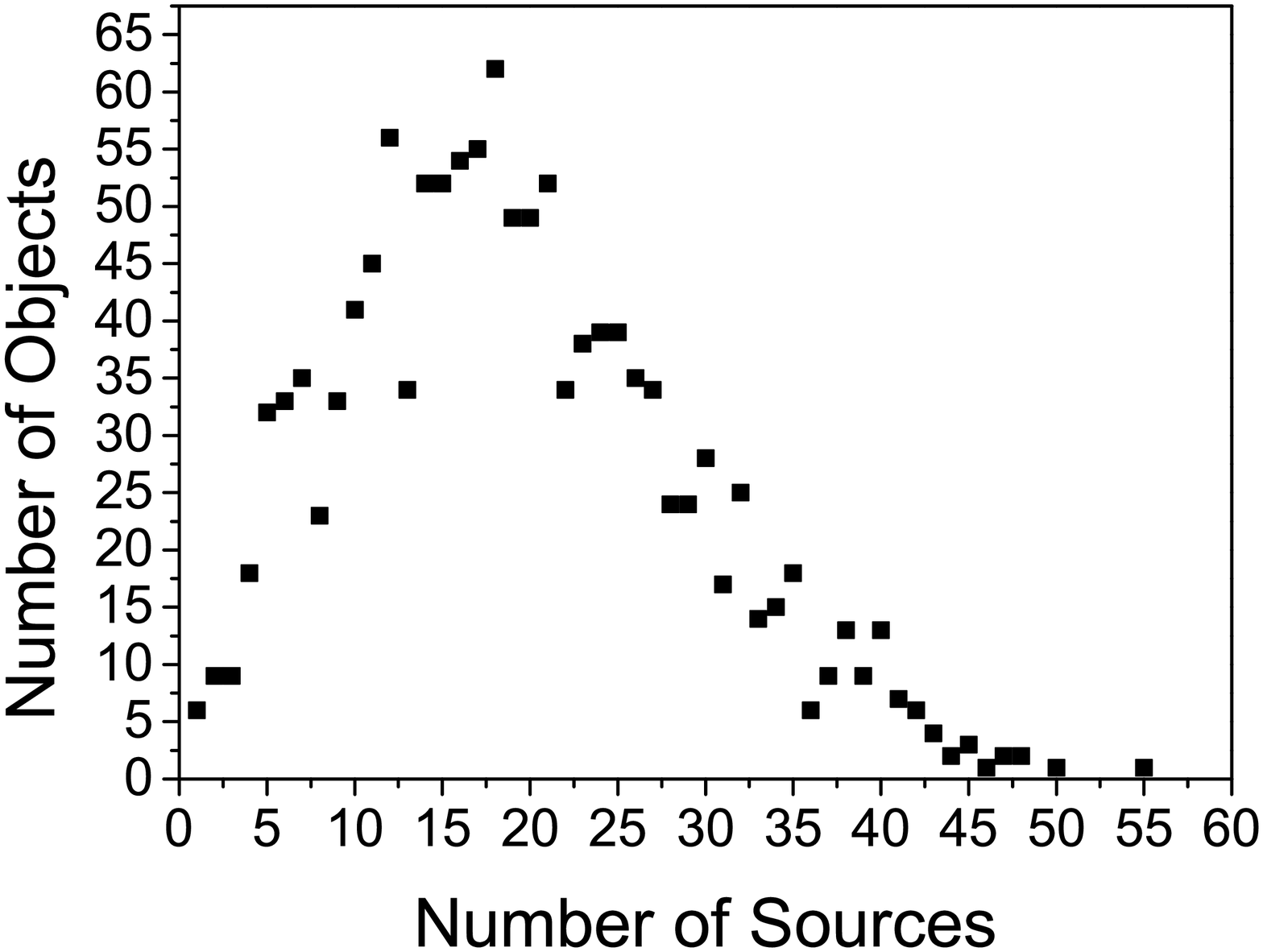}\label{fig:Statistics-a}}
~~
\subfloat[][\scriptsize{Biography Dataset}]{\includegraphics[width=1.7in]{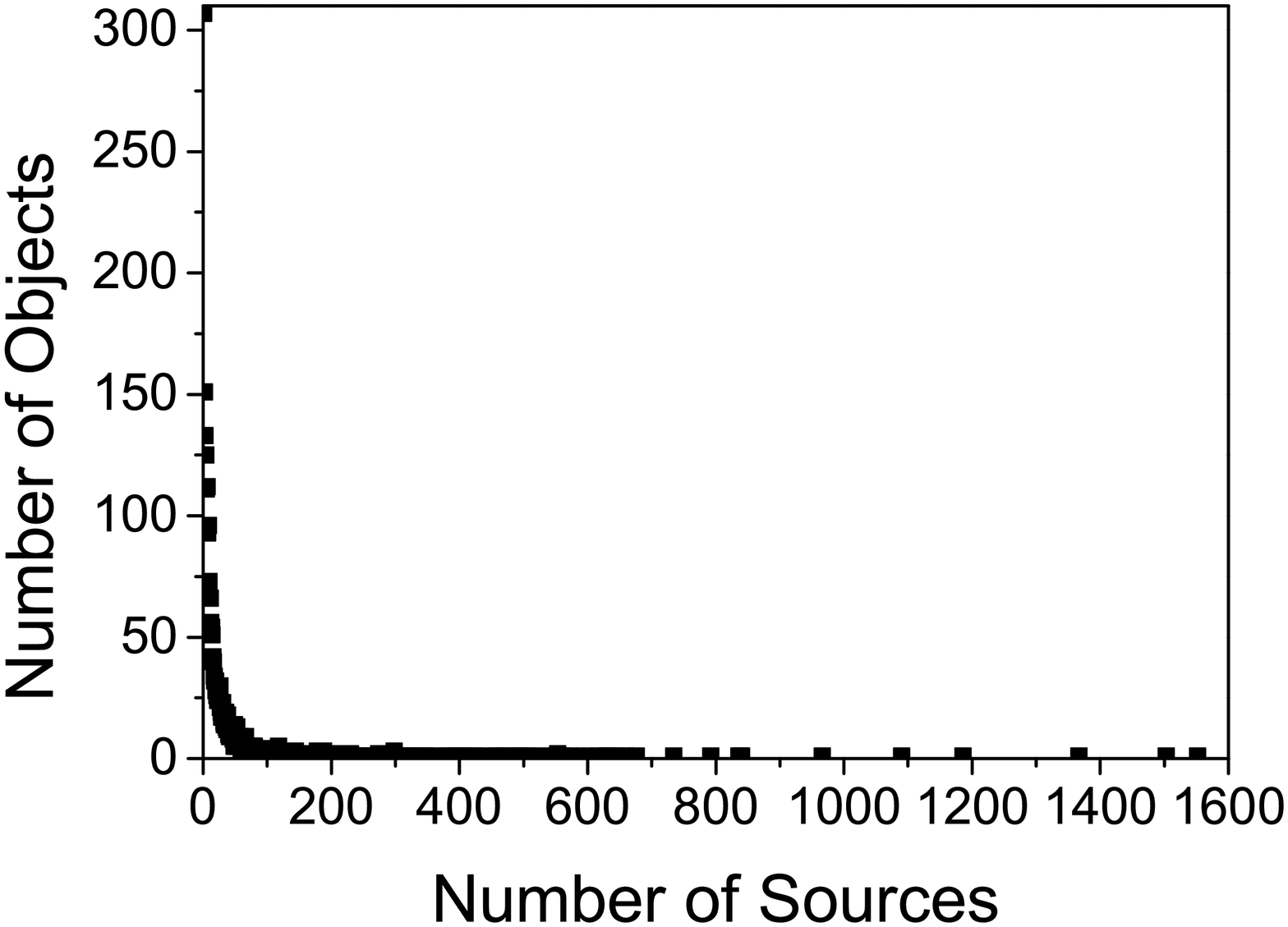}\label{fig:Statistics-b}}
 \vspace{-2mm}
\caption{The number of sources that provide values on objects: different objects are covered by varying numbers of sources.}
  \vspace{-2mm}
\label{fig:Statistics}
\end{figure}

We have investigated the distributions of objects over sources in various real-world datasets. As an example, Fig.~\ref{fig:Statistics-a} and Fig.~\ref{fig:Statistics-b} show the results on the \emph{Book-Author}~\cite{yin2008truth} and \emph{Biography}\footnote{In this paper, we focus on the parent-children relation in the dataset, where \textit{the children of a person} represents a multi-valued object.}~\cite{pasternack2010knowing} datasets, respectively. Each point $(x, y)$ in the figure depicts $y$ objects are covered by $x$ sources in the corresponding dataset. We observe an apparent long-tail phenomenon from the distributions of Biography dataset (contains $2,579$ objects), which indicates that very few objects are referenced by a large number of sources in the dataset; instead, many objects are covered by very few sources. For the Book-Author dataset, which contains much fewer (around $1,262$) objects and ($624$) sources, the long-tail phenomenon is less evident, but objects are claimed by significantly varying numbers of sources, indicating 
that 
objects are of different occurrences. For example, the book ($id:1558606041$) is covered by $55$ sources, while books ($id:0201608359$) and ($id: 020189551X$) are only covered by one source each.

Intuitively, sources tend to gain more attention from the public if they publish more popular information, and the objects with more occurrences in the sources' claims indicate that they are more popular. Since the size of the potential audience is usually bigger 
for popular objects than for less popular objects, a source will mislead more people if it provides false values on a popular object than on a less popular object. With this consideration, we believe there are different impacts of knowing the true values of different objects. Therefore, we propose to distinguish source reliability by differentiating the \emph{popularity} of objects, to minimize the number of people misguided by false values. In this way, sources that provide false values for popular objects can be penalized heavier and assigned with lower reliability, to discourage them in misguiding the public. Meanwhile, sources providing false values for less popular objects would not be penalized aggressively. Moreover, from the data sufficiency's point of view, popular objects are generally claimed by more sources than the less popular objects, and more evidence can be used for estimating value veracity regarding those objects, leading to more reliable truth estimation. This supports the rationale of assigning more weights to popular objects in the calculation of source reliability, which indirectly helps deliver more accurate estimation.

\subsection{Problem Definition}
\label{subsec:Problem Definition}
A multi-truth discovery problem (i.e., MTD) generally involves five components (Table~\ref{tab:notations} summarizes the notations used in this paper) during its life cycle:

\vspace{2mm}
\noindent{\bf Explicit inputs} include: i) a set of \emph{multi-valued objects}, $\mathcal O$, each of which may have more than one true value to be discovered. The numbers of true value(s) can vary from object to object; ii) a set of \emph{sources}, $\mathcal S$. Each $s \in \mathcal S$ provides potential true values on a subset of objects in $\mathcal O$; iii) \emph{claimed values}, the values provided by any source of $\mathcal S$ on any object of $\mathcal O$. Given a source $s$, we regard the set of values provided by $s$ on object $o$ as \emph{positive claims}, denoted as $\mathcal{V}_{s_o}$. As an example, source $s_2$ in Table~\ref{tab:Raw_data} claims two values on the cast of movie ``Harry Potter'', denoted by $\mathcal{V}_{{s_2}_o}$=\{``Emma Watson'', ``Rupert Grint''\}.

\begin{table}[h]
\begin{center}
\begin{scriptsize}
 
\caption{\label{tab:Raw_data}An illustrative example: three sources provide values on the cast of movie ``Harry Potter''}
 
\begin{tabular}{|l|l|l|}
\hline
Sources & Positive Claims & Negative claims \\ \hline
$s_1$                           & \begin{tabular}[c]{@{}l@{}}Daniel Radcliffe, Emma Watson, \\ Rupert Grint\end{tabular}                         & Jonny Depp   \\ \hline
$s_2$                           & Emma Watson, Rupert Grint                               &  \begin{tabular}[c]{@{}l@{}}Daniel Radcliffe, \\ Jonny Depp\end{tabular} \\ \hline
$s_3$                           & \begin{tabular}[c]{@{}l@{}}Daniel Radcliffe, Emma Watson, \\ Jonny Depp\end{tabular} & Rupert Grint \\ \hline
\end{tabular}
\end{scriptsize}
\end{center}
\end{table}

\vspace{2mm}
\noindent{\bf Implicit inputs} are derived from the explicit inputs and include: i) the complete set of  values provided by all sources on any object $o$, denoted as $\mathcal{U}_o$. For example, based on the values the three sources provide on $o$ in Table~\ref{tab:Raw_data}, we can obtain $\mathcal{U}_o$=\{``Daniel Radcliffe'', ``Emma Watson'', ``Rupert Grint'', ``Jonny Depp''\}; ii) by incorporating the \emph{mutual exclusion assumption}, given an object $o$, a source $s$ that make positive claims $\mathcal{V}_{s_o}$ is believed to implicitly disclaim all the other values on $o$. We denote the set of values disclaimed by $s$ as $\mathcal{\tilde V}_{s_o}$ (i.e., \emph{negative claims} provided by $s$ on $o$), which is calculated by $\mathcal{U}_o-\mathcal{V}_{s_o}$. In Table~\ref{tab:Raw_data}, the negative claims of $s_2$ on $o$ is denoted as $\mathcal{\tilde V}_{{s_2}_o}$=\{``Daniel Radcliffe'', ``Jonny Depp''\}.

\vspace{2mm}
\noindent{\bf Intermediate variables} are generated and updated during the iterative truth discovery procedure. They include: i) \emph{source reliability}, which reflects the capability of each source providing true values; ii) \emph{confidence scores}, which reflects the confidence on a value's being true or false. In this paper, we differentiate the false positives and false negatives made by sources by modeling two aspects of source reliability, namely \emph{positive precision} (denoted as ${\tau}(s)$), i.e., the probability of the positive claims of a source being true, and \emph{negative precision} (denoted as ${\tilde\tau}(s)$), i.e., the probability of the negative claims of a source being false. Accordingly, we estimate both the confidence scores of a value $v$ being true (i.e., $\mathcal{C}_v$) and false ($\mathcal{C}_{\tilde v}$). 

\vspace{2mm}
\noindent{\bf Outputs} are the \emph{identified truth} for each object $o \in \mathcal O$ and are denoted as ${\mathcal{V}_o}^*$.

\vspace{2mm}
\noindent{\bf Ground truth}
are the factual true values for each object $o \in \mathcal O$, denoted as ${\mathcal{V}_o}^g$, which are used to be compared with the results of truth discovery methods.

Based on the above analysis, we formally define the multi-truth discovery problem as follows:

\vspace{2mm}
\begin{definition}{\textbf{Multi-Truth Discovery Problem (MTD)}} Given a set of multi-valued objects ($\mathcal O$) and a set of sources ($\mathcal S$) that provide conflicting values $\mathcal V$. The goal of MTD is to identify a set of true values (${\mathcal{V}_o}^*$) from $\mathcal V$ for each object $o$, satisfying that ${\mathcal{V}_o}^*$ is as close to the ground truth ${\mathcal{V}_o}^g$ as possible. A truth discovery process often proceeds along with the estimation of the reliability of sources, i.e., positive precision (${\tau}(s)$) and negative precision (${\tilde\tau}(s)$). The perfect truth discovery results satisfy ${\mathcal{V}_o}^*={\mathcal{V}_o}^g$.
$\square$
\end{definition}

\begin{table*}[!th]
\centering
 
\caption{Notations used in the paper\label{tab:notations}}
 
{
\begin{scriptsize}
\begin{tabular}{|c|l|}
\hline
Notation & Explanation\\\hline\hline
$o$, $\mathcal O$  & An object (resp., Set of all objects)\\
$s$, $\mathcal S$ & A source (resp., Set of all sources)\\
$v$, $\mathcal V$ & A claimed value (resp., Set of all claimed values)\\
${\mathcal{V}_o}^*$ & Identified truth for $o$\\
${\mathcal{V}_o}^g$ & Ground truth for $o$\\\hline
$\mathcal{S}_o$ & Set of sources provide values on $o$\\
$\mathcal{S}_v$, $\mathcal{S}_{\tilde v}$ & Set of sources claim (resp., disclaim) $v$ on $o$\\
$\mathcal{O}_s$ & Set of objects covered by $s$\\
$Cov(s)$ & The coverage of $s$\\
$\mathcal{V}_{s_o}$, $\mathcal{\tilde V}_{s_o}$ & Set of positive (resp., negative) claims provided by $s$ on $o$ \\
$\mathcal{U}_o$ & Set of all claimed values on $o$\\\hline
${\tau}(s)$, ${\tilde\tau}(s)$ & positive (resp., negative) precision of $s$\\
$\mathcal{C}_v$, $\mathcal{C}_{\tilde v}$ & The confidence score of $v$ being true (resp., false)\\
$\mathcal{A}(s_1, s_2)$, $\mathcal{\tilde A}(s_1, s_2)$ & Endorsement degree from $s_1$ to $s_2$ on positive (resp., negative) claims\\
$A_o(s_1, s_2)$, $\tilde A_o(s_1, s_2)$ & Agreement between the positive (resp., negative) claims of $s_1$ and $s_2$ on $o$\\
$\mathcal{P}_o$ & The popularity degree of $o$\\
$\mathcal{D}(s,o)$, $\mathcal{\tilde D}(s,o)$ & The dependence score of $s$ providing positive (resp., negative) claims on $o$\\
$\omega(s_1 \to s_2)$, $\tilde\omega(s_1 \to s_2)$ & The weight of edge from $s_1$ to $s_2$ in $\pm$supportive agreement graph\\
${\omega_{c_o}}(s_1 \to s_2)$, ${\tilde\omega_{c_o}}(s_1 \to s_2)$ & The weight of edge from $s_1$ to $s_2$ in $\pm$malicious agreement graph of $o$\\\hline
\end{tabular}
\end{scriptsize}}
\end{table*}

\subsection{Agreement as Hint}
\label{subsec:Agreement as Hint}
For multi-valued objects, sources may provide totally different, the same, or overlapping sets of values from one another. Given an object, we define the common values claimed by two sources on the object as inter-source \emph{agreement}. Based on the mutual exclusion, we consider two-sided inter-source agreements, where the +\emph{agreement} (resp., --\emph{agreement}) is the agreement between two sources on positive (resp., negative) claims. Intuitively, the agreement among sources indicates 
an {\em endorsement}. If the positive (resp., negative) claims of a source are agreed/endorsed by many other sources, this source may have a high positive (resp., negative) precision and is called an \emph{authoritative source}.

Suppose ${\mathcal{V}_o}^g$ is the ground truth of an object $o$, $\mathcal{U}_o$ is the set of all claimed values on $o$, we denote by $\mathcal{U}_o-{\mathcal{V}_o}^g$ the set of false values of $o$. For the simplicity of presentation, in the following, we use $T$, $U$, and $F$ to represent ${\mathcal{V}_o}^g$, $\mathcal{U}_o$, and $\mathcal{U}_o-{\mathcal{V}_o}^g$ in this section. For any two sources $s_1$ and $s_2$, the +agreement between them on an object $o$ is calculated as:
\begin{equation}
\label{equa:1}
A_o(s_1, s_2)=\mathcal{V}_{{s_1}_o}  \cap \mathcal{V}_{{s_2}_o}
\end{equation}
Suppose $s_1$ and $s_2$ each selects a true value from $T$ independently. We denote their selected values as $t_1$ and $t_2$, respectively. The probability of $t_1=t_2$, $P_{A_o}(t_1, t_2)$, is calculated as follows\footnote{Note that this probability is based on the prior knowledge that $s_1$ and $s_2$ each provides a true value, which is different from the probability of two sources $s_1$ and $s_2$ independently providing the same true value.}:
\begin{equation}
\label{equa:2}
P_{A_o}(t_1, t_2)=\frac{1}{\left | T \right |}
\end{equation}

Similarly, let $f_1$ and $f_2$ be the two values independently selected by $s_1$ and $s_2$ from $F$, and $P_{A_o}(f_1, f_2)$ be the probability of $s_1$ and $s_2$ providing the same false value (i.e., $f_1=f_2$). we have:
\begin{equation}
\label{equa:3}
P_{A_o}(f_1, f_2)=\frac{1}{\left | F \right |}
\end{equation}

In reality, an object usually has a small truth set and random false values, i.e., $|T| \ll |U|$. Applying this to Equation~\eqref{equa:2} and Equation~\eqref{equa:3}, we get:
\begin{equation}
\label{equa:4}
P_{A_o}(f_1, f_2)\ll P_{A_o}(t_1, t_2)
\end{equation}
Usually, the values claimed by sources contain a fraction of values from each of $T$ and $F$. According to Equation~\eqref{equa:4}, positive claims from $T$ are more likely to agree with each other than negative claims from $F$ to agree with each other. This implies that the more true values a source claims, the more likely the other sources would agree with its claimed values.
Inversely, if a source shows a high degree of agreement with other sources regarding its claimed values, the values claimed by this source would have a higher probability to be true, and this source would have a higher positive precision.

Similarly, the --agreement between any two sources $s_1$ and $s_2$ on an object $o$ is calculated as:
\begin{equation}
\label{equa:5}
\tilde A_o(s_1, s_2)=\mathcal{\tilde V}_{{s_1}_o}  \cap \mathcal{\tilde V}_{{s_2}_o}=U-(\mathcal{V}_{{s_1}_o}  \cup \mathcal{V}_{{s_2}_o})
\end{equation}

Let $\tilde A_o(s_1, s_2) \cap T$ be the agreement of true values and $\tilde A_o(s_1, s_2) \cap F$ be the agreement of false values, satisfying $|\mathcal{V}_{{s_1}_o}| \ll |U|$, $|\mathcal{V}_{{s_2}_o}| \ll |U|$, $|T| \ll |U|$. It can be proved that $|\tilde A_o(s_1, s_2) \cap T| \ll |\tilde A_o(s_1, s_2) \cap F|$. Therefore, it is more likely for sources to agree with each other on false values than true values with respect to their negative claims. This implies that the more false values a source disclaims, the more likely the other sources would agree with its negative claims; inversely, if a source shows a high degree of agreement with the other sources on its negative claims, the values disclaimed by this sources would have higher probabilities to be false, and this source would have a higher negative precision.

\section{The SmartMTD Approach}
\label{sec:Approach}
In reality, sources might not only support one another by providing the same true claims but also
may maliciously copy from others to provide the same false claims, which sometimes mislead the audience. 
Therefore, we identify two types of source relations. Specifically, sharing the same true values means one source implicitly supports/endorses the other source, indicating a \emph{supportive relation} between two sources. 
We define the common values between these two sources as a \emph{supportive agreement}. 
Based on the analysis in Section~\ref{subsec:Agreement as Hint}, we can measure source reliability by quantifying sources' supportive agreement. Even though one source can copy from one another, we consider this type of copying relations as benignant. On the contrary, sharing the same false values is typically a rare event when the sources are fully independent. If two sources share a significant amount of false values, they are likely to copy from each other, indicating a \emph{copying relation} between them. We define these common false values as a \emph{malicious agreement} and quantify the dependence degrees of sources because neglecting the existence of deliberate copying of false values would impair the accuracy of source reliability estimation. 

Additionally, previous research efforts do not differentiate the popularity of different objects. However, in reality, the impact of knowing the true values of different objects might differ. For example, between the email addresses and the children of a famous researcher, the email addresses are apparently more popular and have bigger impacts as other researchers or students doing research in the same areas often need to contact him/her. Taking object popularity into consideration could better model the real-world truth discovery and therefore lead to more accurate results. 

Based on the above observations, we propose a graph-based model, called \emph{SmartMTD}, which incorporates two implications, i.e., two-sided source relations and object popularity, to solve the MTD problem.

\subsection{The Graph-Based Model}
\label{subsec:The Graph-Based Model}
SmartMTD applies the following principle for truth discovery~\cite{li2015survey}: sources providing more true values are assigned with high reliability; meanwhile, values provided by high-quality sources are more likely to be selected as true values. 

To measure the two aspects of source reliability, we construct two fully connected weighted graphs, namely $\pm$\emph{sup-portive agreement graph}s, based on modeling the two-sided supportive relations among sources. In both graphs, the vertices denote sources, each directed edge represents that one source agrees with the other source, and the weight on each edge depicts to what extent one source endorses the other source. We define $\mathcal{A}(s_1, s_2)$ (resp., $\mathcal{\tilde A}(s_1, s_2)$) as the \emph{endorsement degree} from $s_1$ to $s_2$ on positive (resp., negative) claims, representing the rate at which $s_2$ is endorsed by $s_1$ on the value's being true (resp., false). We measure the endorsement by quantifying the supportive agreement and taking the \emph{copying relations} among sources and \emph{object popularity} into account. We will introduce the methods for malicious agreement detection and object popularity quantification in the following two subsections, respectively.

\vspace{2mm}
\noindent{\bf +Supportive Agreement Graph}. We first formalize the endorsement between two sources based on their common positive claims as follows:
\begin{equation}
\label{equa:6}
\mathcal{A}(s_1, s_2)=\sum_{o \in \mathcal{O}_{s_1}  \cap \mathcal{O}_{s_2}}{\frac{|A_o(s_1, s_2)|}{|\mathcal{V}_{{s_2}_o}|}}\cdot (1-\prod_{v \in A_o(s_1, s_2)}{\mathcal{C}_{\tilde v}})\cdot\mathcal{P}_o \cdot (1- \mathcal{D}(s_1,o))
\end{equation}
where $\mathcal{D}(s_1,o)$ denotes the dependence score of $s_1$ providing positive claims on $o$ (defined in Section~\ref{subsec:Copy}), and $\mathcal{P}_o$ denotes the popularity degree of $o$ (Section~\ref{subsec:popularity}).

We calculate the weight on the edge from $s_1$ to $s_2$ using:
\begin{equation}
\label{equa:7}
\omega(s_1 \to s_2)= \beta+(1-\beta)\cdot\frac{\mathcal{A}(s_1, s_2)}{|\mathcal{O}_{s_1}  \cap \mathcal{O}_{s_2}|}
\end{equation}
where $\beta$ is the smoothing factor.  By assigning a small weight to every pair of vertices, we actually add a ``\emph{smoothing link}'' to each possible edge of the graphs. This measure guarantees the graph is fully connected and the calculation of source positive precision can converge. We set $\beta$ at $0.1$ for our experiments. Empirical studies like Gleich et al.~\cite{Gleich2010surfer} may help more accurate estimation. Finally, we normalize the weights of out-links from every vertex by dividing the edge weights by sum of the out-going edge weights from the vertex. This normalization allows us to interpret the edge weights as transition probabilities for the random walk computation. 

\vspace{2mm}
\noindent{\bf --Supportive Agreement Graph}. We construct the --supportive agreement graph in a similar way by applying the following equations:
\begin{equation}
\label{equa:8}
\mathcal{\tilde A}(s_1, s_2)=\sum_{o \in \mathcal{O}_{s_1}  \cap \mathcal{O}_{s_2}}{\frac{|\tilde A_o(s_1, s_2)|}{|\mathcal{\tilde V}_{{s_2}_o}|}}\cdot (1-\prod_{v \in {\tilde A}_o(s_1, s_2)}{\mathcal{C}_v})\cdot\mathcal{P}_o \cdot (1-\mathcal{\tilde D}(s_1,o))
\end{equation}
\begin{equation}
\label{equa:9}
\tilde\omega(s_1 \to s_2)= \beta+(1-\beta)\cdot\frac{\mathcal{\tilde A}(s_1, s_2)}{|\mathcal{O}_{s_1}  \cap \mathcal{O}_{s_2}|}
\end{equation}

Specifically, we adopt the \emph{Fixed Point Computation Model} (FPC) to calculate the positive precision and negative precision of each source. FPC captures the transitive propagation of source trustworthiness through agreement links based on the above-constructed two graphs~\cite{brin1998pagerank}. In particular, we refer to each graph as a Markov chain, with vertices regarded as states, and the weights on the edges as the probabilities of transition between states. We calculate the asymptotic stationary visit probabilities of the Markov random walk. As the sum of all visit probabilities equals to $1$, they cannot reflect the real source precision. To resolve this issue, we set the positive precision (resp., negative precision) of the source with the highest visit probability in the +supportive agreement graph (resp., --supportive agreement graph) as $pp_{max}$ (resp., $np_{max}$), and calculate the \emph{normalization rate} by dividing the precision by the corresponding visit probability. Then, the visit probabilities of all sources can be normalized as precision, denoted as $\tau(s)$ and $\tilde\tau(s)$, by multiplying the normalization rate. The computed precision captures the following characteristics:

\begin{itemize}
\item Vertices with more input edges have higher precision since those sources are endorsed by a large number of sources and should be more trustworthy. Here, we neglect the smoothing links. If there is no common value between two sources, there is no link between them in the graphs. Endorsement from a source with more input edges should be more trusted than that from other sources. Since an authoritative source is likely to be more trustworthy, the source endorsed by an authoritative source is also more likely to be trustworthy.
\item The endorsement on values with higher probability to be true (resp., false) in the +supportive agreement graph should be more (resp., less) respected. Meanwhile, the endorsement of values with higher probability to be false (resp., true) in the --supportive agreement graph should be more (resp., less) respected.
\item Endorsement from a source on popular objects should be highlighted since popular objects have a bigger impact on the public and false values of the popular object can lead to worse consequences. Meanwhile, the endorsement provided by a malicious copier should be penalized.
\end{itemize}
To jointly determine value veracity from source reliability, we assume each source of $\mathcal{S}_o$ contributes a smart vote to each potential value of $o$. In particular, if a source provides $v$ as a positive claim, then it casts a vote proportional to ${\tau}'(s)$ for it; in contrast, if a source disclaims $v$, then it casts a vote proportional to $(1-{\tau}'(s))$ for it. Therefore, we compute the confidence score of each value $v$ being true and false as follows:
\begin{equation}
\label{equa:10}
\mathcal{C}_v=\frac{\sum_{s \in \mathcal{S}_v}{{\tau}'(s)}+\sum_{s \in \mathcal{S}_{\tilde v}}{(1-{\tilde\tau}'(s))}}{|\mathcal{S}_o|}
\end{equation}
\begin{equation}
\label{equa:11}
\mathcal{C}_{\tilde v}=\frac{\sum_{s \in \mathcal{S}_v}{(1-{\tau}'(s))}+\sum_{s \in \mathcal{S}_{\tilde v}}{{\tilde\tau}'(s)}}{|\mathcal{S}_o|}
\end{equation}

\subsection{Detecting Malicious Agreement}
\label{subsec:Copy}
Copying relations among sources in real world can be complex. For example, a copier may copy all values or partial values from a source; a source may transitively copy from another source; and one source can be copied by multiple sources. To model the malicious agreement among sources globally, we construct $\pm$\emph{malicious agreement graphs} for sources that provide values on each object $o$, i.e., $\mathcal{S}_o$. Similar to the graphs constructed above, each edge of the +malicious (resp., -malicious) agreement graph represents one source maliciously endorsing the other on the positive (resp., negative) claims regarding an object with a quantified endorsement degree, denoted as ${\omega_{c_o}}(s_1 \to s_2)$ (resp., ${\tilde\omega_{c_o}}(s_1 \to s_2)$):
\begin{equation}
\label{equa:12}
{\omega_{c_o}}(s_1 \to s_2)= \beta+(1-\beta)\cdot\frac{|A_o(s_1, s_2)|}{|\mathcal{V}_{{s_2}_o}|} \cdot (1-\prod_{v \in A_o(s_1, s_2)}{\mathcal{C}_v})
\end{equation}
\begin{equation}
\label{equa:13}
{\tilde\omega_{c_o}}(s_1 \to s_2)= \beta+(1-\beta)\cdot\frac{|{\tilde A}_o(s_1, s_2)|}{|\mathcal{\tilde V}_{{s_2}_o}|} \cdot (1-\prod_{v \in {\tilde A}_o(s_1, s_2)}{\mathcal{C}_{\tilde v}})
\end{equation}

Both the FPC random walk computation and normalization are conducted on each graph to obtain the dependence scores for sources that provide positive (resp., negative) claims on an object $o$, denoted as $\mathcal{D}(s,o)$ (resp., $\mathcal{\tilde D}(s,o)$). We set the dependent score of the source with the highest visit probability in the +malicious agreement graph (resp., --malicious agreement graph) as $pc_{max}$ (resp., $nc_{max}$).

The computed dependence scores capture the following characteristics:

\begin{itemize}
\item Vertices with more input edges have a higher value of dependence score since those sources are maliciously endorsed by a larger number of sources. Such sources act as collectors that copy values from other sources.
\item The malicious endorsement on values with lower probabilities to be true (resp., false) in the +malicious agreement graph should be more (resp., less) respected. Meanwhile, the malicious endorsement on values with lower probabilities to be false (resp., true) in the --malicious agreement graph should be more (resp., less) respected as well.
\end{itemize}

\begin{algorithm}[!tb]
\begin{scriptsize}
\KwIn{objects of interest $\mathcal O$, sources $\mathcal S$, and $\mathcal{V}_{s_o}$ the set of positive claims provided by $s$ on $o$.}
\KwOut{${\mathcal{V}_o}^*$ identified truth for each $o \in \mathcal O$.}
\tcp{\scriptsize{Initialization phase}}
Initialize $\delta$, $\beta$, $pp_{max}$, $np_{max}$, $pc_{max}$, $nc_{max}$\\
Initialize $\mathcal{C}_v$, $\mathcal{C}_{\tilde v}$ for each $v\in \mathcal V$, $o\in \mathcal O$\\

\tcp{\scriptsize{Object popularity quantification}}

\ForEach{$o\in \mathcal O$}{
compute $\mathcal{P}_o$  by Equation~\eqref{equa:14},~\eqref{equa:16}
}

\tcp{\scriptsize{Iteration phase}}
\Repeat{convergence}{
\tcp{\scriptsize{Malicious agreement detection}}

\ForEach{$o\in \mathcal O$}{
construct $\pm$malicious agreement graphs by quantifying the weights of each edge by Equation~\eqref{equa:12},~\eqref{equa:13}\\
derive $\mathcal{D}(s,o)$, $\mathcal{\tilde D}(s,o)$ by applying random walk and normalization steps\\
}

\tcp{\scriptsize{$\pm$Source Reliability computation}}

construct $\pm$supportive agreement graphs by quantifying the weights of each edge by Equation~\eqref{equa:6},~\eqref{equa:7},\eqref{equa:8},~\eqref{equa:9}\\
derive ${\tau}'(s)$, ${\tilde\tau}'(s)$  by applying random walk and normalization steps\\

\tcp{\scriptsize{Value confidence score computation}}

\ForEach{$v\in \mathcal V$, $o\in \mathcal O$}{
        compute $\mathcal{C}_v$, $\mathcal{C}_{\tilde v}$ by Equation~\eqref{equa:10},~\eqref{equa:11}\\
    }
}
\Return $\{(o,v)|v\in \mathcal V \wedge \mathcal{C}_v>\mathcal{C}_{\tilde v} \wedge v\in \mathcal{U}_o, o\in \mathcal{O}\}$\\
\caption{The Algorithm of SmartMTD.}
\label{alg:1}
\end{scriptsize}
\end{algorithm}

\subsection{Quantifying Object Popularity}
\label{subsec:popularity}
Intuitively, popular objects tend to be covered by more sources, as sources tend to publish popular information to attract more audiences. Therefore, objects covered by more sources are normally more popular than those covered by fewer sources. We quantify the \emph{popularity} of each object, i.e., $\mathcal{P}_o$, by its occurrence frequency in sources' claims. Specifically, we consider each source casts a vote for the popularity of each object it covers, and the votes from all the sources that claim values on an object jointly determine the object's popularity. We define the coverage of a source $s$, i.e., $Cov(s)$, as the percentage of its covered objects over $\mathcal O$. The votes for the popularity of an object from sources with lower coverage should be more respected than those from the sources with higher coverage, as popular
objects will be more conspicuous in small sources. Formally, we measure the popularity of each object by applying the following equations, which comprehensively incorporates the occurrence of the object and the coverage of each source that provides the object:
\begin{equation}
\label{equa:14}
\mathcal{P}_o^u=\sum_{s \in \mathcal{S}_o}{\frac{1}{Cov(s)}}\\
\end{equation}

 
\begin{equation}
\label{equa:16}
\mathcal{P}_o=\frac{\mathcal{P}_o^u}{\sum_{o' \in \mathcal O}{\mathcal{P}_{o'}^u}}\\
\end{equation}
where $\mathcal{P}_o^u$ is the unnormalized popularity of object $o$.

\subsection{The Algorithm}
\label{subsec:The Algorithm}
Algorithm~\ref{alg:1} shows the procedure of SmartMTD. In the initialization phase, the parameters, including the iteration convergence threshold $\delta$, smoothing factor $\beta$, positive precision $pp_{max}$, negative precision $np_{max}$, the two-sided dependence scores ($pc_{max}$ and $nc_{max}$) of sources with the highest visit probabilities in $\pm$supportive agreement graphs and $\pm$malicious agreement graphs, are initialized with their a priori values (line 1). The confidence scores of each value $v$ being true or false are both initialized by adopting the majority voting in our experiments (in fact, other truth discovery methods can also be applied for this initialization). $\mathcal{C}_{\tilde v}$ is initialized as $1-\mathcal{C}_v$ (line 2).
To start, we count the votes of each individual value of each object, and then normalize those vote counts by dividing them by $|\mathcal{S}_o|$ to represent $\mathcal{C}_v$ for each value;  The \emph{object popularity} (lines 3-4) is calculated directly based on the multi-source data. For each cycle of iteration, the algorithm recalculates the two-sided source dependence scores (lines 6-9), continues to calculate sources' positive precision and negative precision (lines 10-11) based on the two-sided value confidence scores, and computes confidence scores of values (lines 12-13) based on the two-sided source precision. The algorithm examines the difference of \emph{cosine similarity} of the two-sided source precision between two successive iterations against a threshold, $\delta$, to determine its convergence (line 14).

The time complexity of the algorithm is $O(|\mathcal O||\mathcal S|^2+|\mathcal S|^2+|\mathcal V|)$. There are many mature distributed computing tools that can be used for random walk computation to reduce the time complexity. For example, Apache Hama\footnote{https://hama.apache.org/} is a framework for big data analytics, which uses the \emph{Bulk Synchronous Parallel} (BSP) computing model. It provides the \emph{Graph} package for vertex-centric graph computation. 
It should be noted that we can easily extend the \emph{Vertex} class to create a class for realizing parallel random walk computation.

\section{Experiments}
\label{sec:Experiments}
In this section, we report the experimental studies on the comparison of our approach with the state-of-the-art algorithms using real-world datasets, and the impact of the two concerns, namely malicious agreement detection and object popularity quantification.

\begin{table*}[]
\centering
\vspace{-1mm} 
\caption{Comparison of different methods: the best and second best performance values are in bold.}
\vspace{-1mm} 
\label{tab:methods_Comparison}
\begin{scriptsize}
\begin{tabular}{|c|c c c|c c c| c |c c c|c c c| c |}
\hline
\multirow{2}{*}{Method} & \multicolumn{7}{c|}{Book-Author Dataset} & \multicolumn{7}{c|}{Parent-Children Dataset} \\
\cline{2-15}
                 & P  & R  & F1  & WP  & WR  & WF1  &T(s) & P   & R   & F1   & WP   & WR   & WF1 &T(s)  \\
\hline\hline
Voting           & \textbf{0.84}    &  0.63   &  0.72    & \textbf{0.83}     & 0.64     &0.72 &  \textbf{0.07}    & 0.88      & 0.85     &  0.87    &      0.69 & 0.68      & 0.69      & \textbf{0.56 }     \\
Sums             &\textbf{0.84}     & 0.64    &  0.73    & \textbf{0.83}     & 0.64     &      0.72 & 0.85     & \textbf{0.90}      & 0.89     & \textbf{0.90}     &      \textbf{0.88} & 0.86      &  0.87     & 1.13      \\
Avg-Log          & \textbf{0.83}    & 0.60    & 0.70     & \textbf{0.83}     &  0.64     &     0.72 & 0.61     & \textbf{0.90}      & 0.89     & 0.89     &\textbf{0.88}      & 0.86      & 0.87      & \textbf{0.75}      \\
TruthFinder      &\textbf{0.84}     & 0.60    & 0.70     & \textbf{0.83}     & 0.60      &     0.70 &  0.74    & \textbf{0.90}      & 0.89     & \textbf{0.90}     &\textbf{0.88}   &  0.85     &  0.86     &1.24       \\
2-Estimates      &0.81     & 0.70    &  0.75    & 0.80     & 0.68      &     0.74 &  \textbf{0.38}    &  \textbf{0.91}     &  0.89    & \textbf{0.90}     &  \textbf{ 0.88}    &  0.86     &  0.87     &  1.34     \\\hline
LTM              & 0.82    &0.65     & 0.73     &\textbf{0.82}      &  0.62     &     0.71 &  0.98    & 0.87      &  0.90    &  0.88    &0.86       &0.89       &0.87       & 0.99      \\
MBM              & \textbf{0.83}    & \textbf{0.74}    & \textbf{0.78}     &\textbf{0.82}      &\textbf{0.71}      &     \textbf{0.76} & 0.67     &\textbf{0.90}       & \textbf{ 0.92}    &  \textbf{0.91}    & 0.87      &  \textbf{0.90}     & \textbf{0.88}      & 2.17      \\
MTD-hrd        &\textbf{0.83}   &0.58    &  0.68    & \textbf{0.82}     & 0.59      &0.69 &0.72 & \textbf{0.90}      & 0.90    &  \textbf{0.90}    & 0.87 & 0.89    &  \textbf{0.88}     & 1.37     \\\hline
SmartMTD        &\textbf{0.83}   &\textbf{0.75}     &  \textbf{0.79}    & \textbf{0.83}     & \textbf{0.78}      &\textbf{0.80} &0.43 & \textbf{0.90}      & \textbf{0.93}     &  \textbf{0.91}    & \textbf{0.93}      & \textbf{0.92 }     &  \textbf{0.93}     & 0.92     \\
\hline
\end{tabular}
\end{scriptsize}
\end{table*}
\subsection{Experimental Setup}
\label{subsec:Experimental Setup}

\subsubsection{The Datasets}
We used two real-world datasets in our experiments. Each object in the both datasets may contain multiple true values. 

\vspace{2mm}
\noindent{\bf{Book-Author dataset}}~\cite{yin2008truth} contains $33,971$ book-author records from \emph{www.abebooks.com}. These records were collected from numerous book websites (i.e., sources). Each record represents a store's positive claims on the author name(s) of a book (i.e., objects). We refined the dataset by removing the invalid and duplicated records, and excluding the records with only minor conflicts to make the problem more challenging---otherwise, even a straightforward method could yield competitive results. We finally obtained $13,659$ distinctive claims, $624$ websites providing values of author name(s) for $677$ books. Each book has on average $3$ authors. The ground truth provided by the original dataset was used as the gold standard.

\vspace{2mm}
\noindent{\bf{Parent-Children dataset}} was prepared by extracting parent-children relations from the \emph{Biography dataset}~\cite{pasternack2010knowing}. We obtained $227,583$ claims on $2,579$ people's children information (i.e., objects) edited by $54,764$ users (i.e., sources). We further removed duplicate and minorly conflicting records in this dataset for more effective comparison. In the resulting dataset, each person has on average $2.48$ children. We used the latest editing records as the gold standard.

\subsubsection{Baseline Methods}
We compared SmartMTD with two types of baselines. The first type of baselines consists of methods under the single-truth assumption. We chose several typical and competitive methods for comparison but excluded the methods that are inapplicable to the MTD. For example, the method in~\cite{pasternack2010knowing} requires the normalization of the veracity scores of values, which is infeasible for MTD; the methods in~\cite{zhao2012probabilistic,li2014resolving} focus on handling heterogeneous data; and the method in~\cite{li2014confidence} is designed for continuous data, while our approach is designed for categorical data.
\begin{itemize}
\item \emph{Voting}: this method regards a value set as true if the proportion of sources that claim the set is higher than any other value sets.
\item \emph{Sums~\cite{kleinberg1999authoritative}, Average-Log~\cite{pasternack2010knowing}}: these two methods are modified to incorporate mutual exclusion. They compute the total reliability of all sources that claim and disclaim a value separately and regards a value as true if the former sum is bigger than the latter on the value.

\item \emph{TruthFinder}~\cite{yin2008truth}: we used the original version of this method, which estimates \emph{trustworthiness of source} and \emph{confidence of fact} alternately from each other by additionally considering the \emph{influences between facts}. 

\item \emph{2-Estimates}~\cite{galland2010corroborating}: the original version of this method is applied, which also adopts mutual exclusion.
\end{itemize}

The second type of baselines consists of three existing MTD methods:

\begin{itemize}
\item \emph{LTM}~\cite{zhao2012bayesian}: the \emph{Latent Truth Model}, which models two different aspects of source quality in a generative model to tackle multi-valued objects.

\item \emph{MBM}~\cite{wang2015integrated}: the \emph{Multi-truth Bayesian Model}, which incorporates a new mutual exclusion definition and finer-grained copy detection techniques in a Bayesian framework.

\item \emph{MTD-hrd}~\cite{wang2016implications}: a model designed for \emph{Multi-Truth Discovery}, which incorporates two implications, namely the calibration of imbalanced positive/negative claim distributions and the consideration of the implication of values' co-occurrence in the same claims, to improve the probabilistic approach.
\end{itemize}

To ensure the fair comparison, 
we used the same stop criterion for all the iterative methods to determine their convergence. For our approach, we simply used the default parameter settings for both datasets. Intuitively, sources tend to provide values that they are sure to be true and omit uncertain values, while copiers are likely to copy those explicitly claimed values from other sources. Therefore, we set $pp_{max}$ as $1$, $np_{max}$ as $0.9$, $pc_{max}$ as $1$, and $nc_{max}$ as $0.8$. We also studied the impact of different concerns on the performance of our approach (to be detailed in Section~\ref{subsec:Impact of Different Concerns}).

\subsubsection{Evaluation Metrics}
We implemented all the above methods in Python 3.4.0 and ran experiments on a 64-bit Windows 10 Pro. PC with an Intel Core i7-5600 processor and 16GB RAM. We ran each method 
10 times 
and used three types of evaluation metrics to evaluate their average performance.
\begin{figure*}
\centering
\subfloat[][Precision, Recall and Execution Time]{\includegraphics[trim=0 0 0 0, clip=true, height=1.5in]{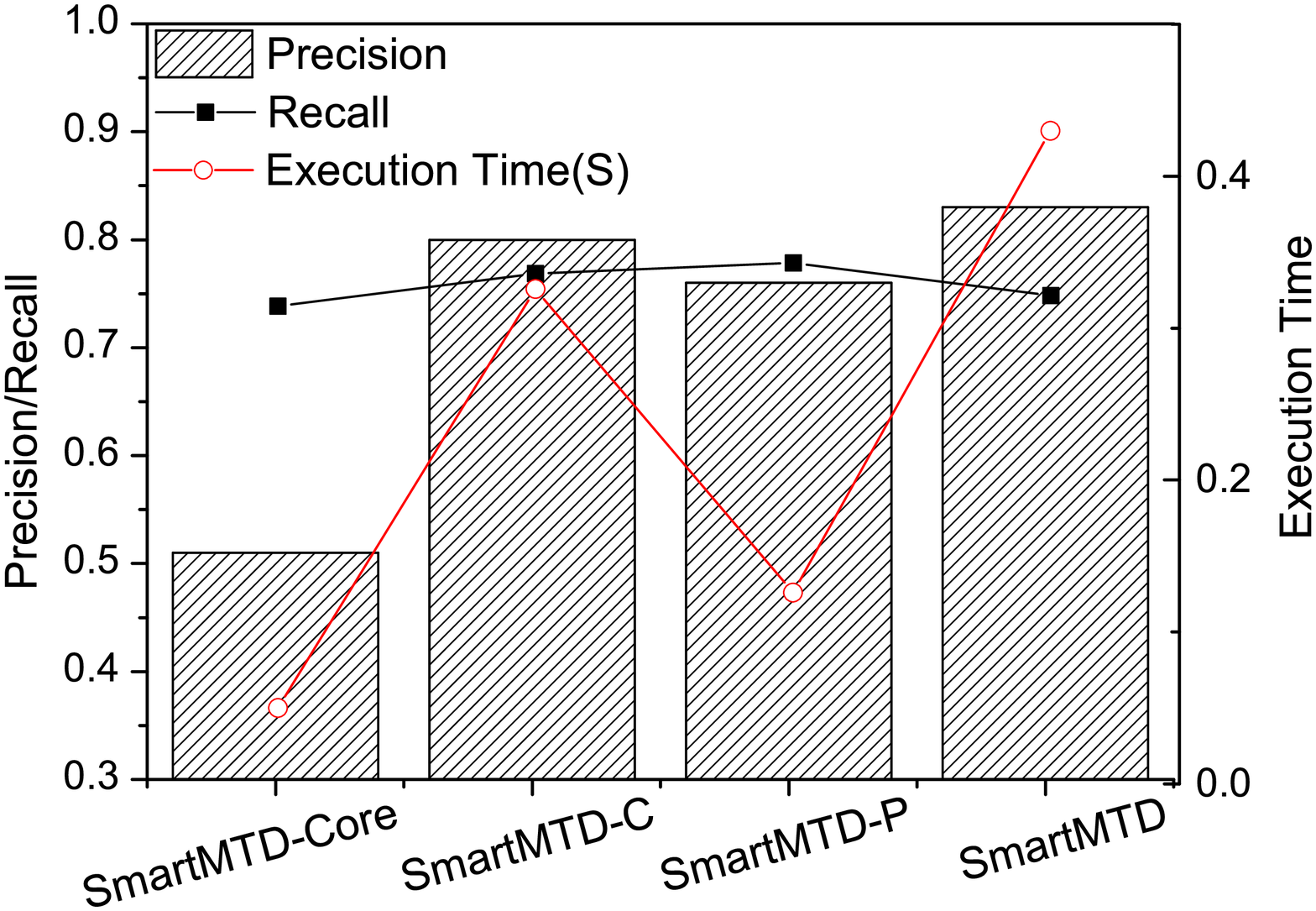}\label{fig:impact-a}}
\hfill
\subfloat[][Weighted Precision and Recall]{\includegraphics[trim=0 0 0 0, clip=true, height=1.5in]{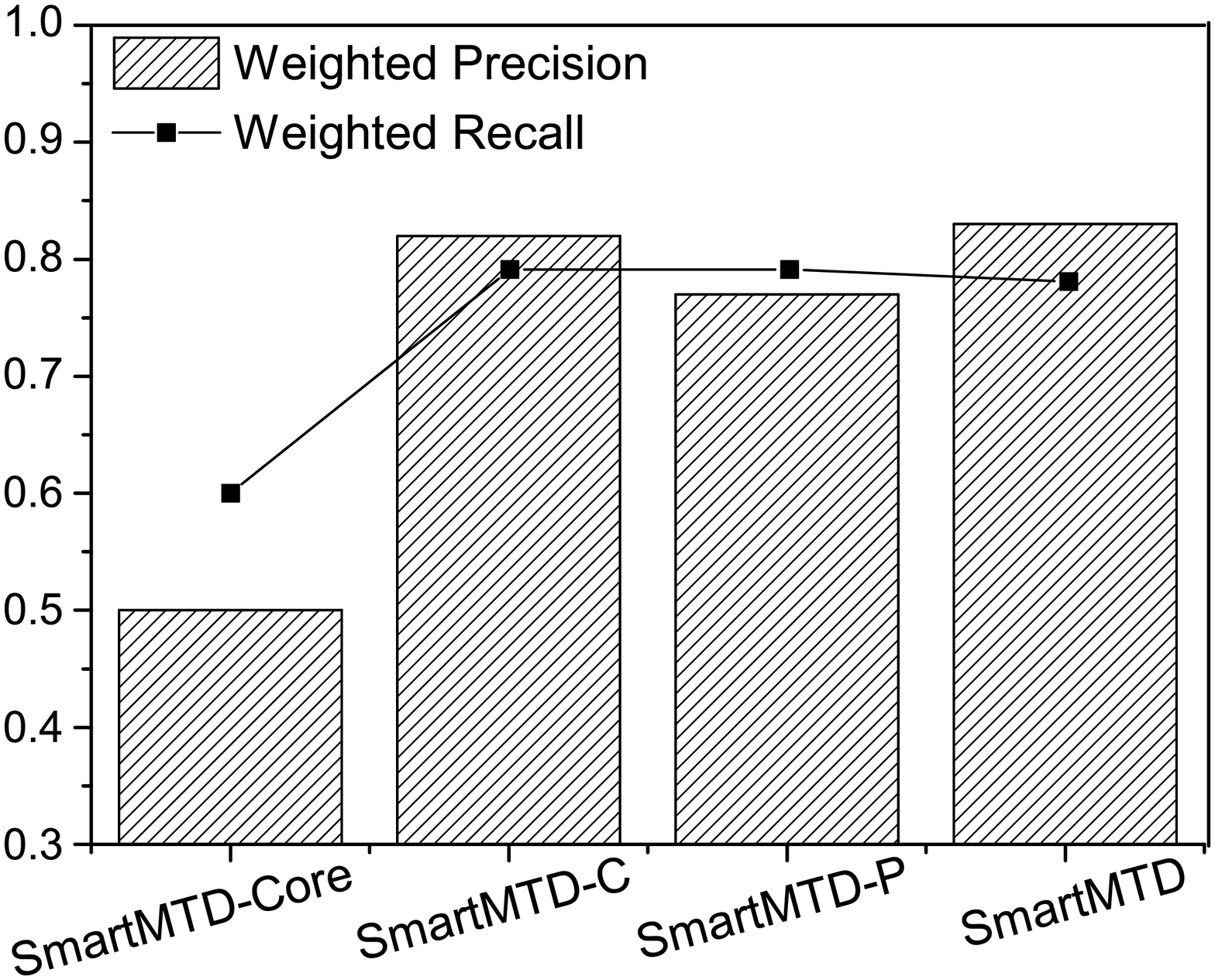}\label{fig:impact-b}}
\hfill
\subfloat[][F$_1$ Score and Weighted F$_1$ Score]{\includegraphics[trim=0 0 0 0, clip=true, height=1.5in]{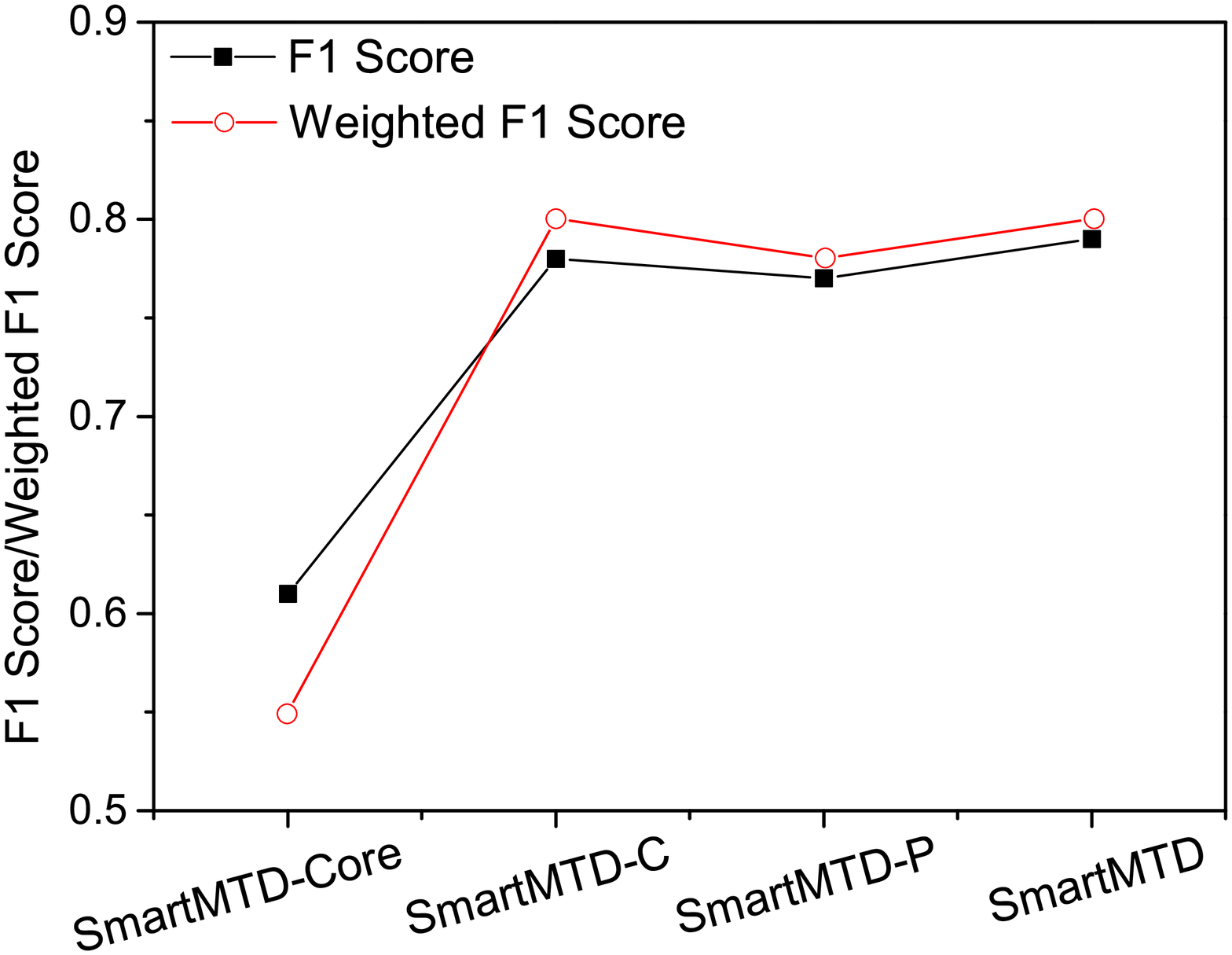}\label{fig:impact-c}}
\vspace{-2mm} 
\caption{Performance comparison of different variants of SmartMTD on Book-Author dataset.}
 \vspace{-2mm} 
\label{fig:impact}
\end{figure*}

\vspace{2mm}
\noindent{\bf{Traditional accuracy metrics. }}\emph{Precision} and \emph{recall} are two commonly used performance measures for evaluating the accuracy of truth discovery methods. We additionally used F$_1$ score as an overall metric as neither precision nor recall could represent the accuracy independently. 
\begin{itemize}
\item \emph{Precision}: the average proportion of predicted actual true values in the set of all returned values on all the objects of a certain truth discovery method:
\begin{equation}
\label{equa:17}
precision=\frac{1}{K|\mathcal{O}|}\sum_{k=1}^{K}{\sum_{n=1}^{|\mathcal{O}|}{\frac{|{\mathcal{V}_o}^{*(k)} \cap {\mathcal{V}_o}^g|}{|{\mathcal{V}_o}^{*(k)}|}}}
\end{equation}
where ${\mathcal{V}_o}^{*(k)}$ is the set of true values identified by $k$-th run of the method for object $o$.
\item \emph{Recall}: the average proportion of predicted actual true values in the set of ground true values on all the objects of a certain truth discovery method:
\begin{equation}
\label{equa:18}
recall=\frac{1}{K|\mathcal{O}|}\sum_{k=1}^{K}{\sum_{n=1}^{|\mathcal{O}|}{\frac{|{\mathcal{V}_o}^{*(k)} \cap {\mathcal{V}_o}^g|}{|{\mathcal{V}_o}^g|}}}
\end{equation}
\item \emph{F$_1$ score}: the harmonious mean of precision and recall, computed as:
\begin{equation}
\label{equa:19}
F_1 \, score=2\cdot\frac{precision\cdot recall}{precision+recall}
\end{equation}
\end{itemize}

\vspace{2mm}
\noindent{\bf{Efficiency metrics. }}\emph{Execution time} was used for efficiency comparison. It is measured by applying the following equation, where $T^{(k)}$ is the execution time of the $k$-th run of a method.
\begin{equation}
\label{equa:20}
execution \, time=\frac{1}{K}\sum_{k=1}^{K}{T^{(k)}}
\end{equation}

\vspace{2mm}
\noindent{\bf{Object popularity weighted accuracy metrics. }}Since we introduce a new concept of object popularity, to measure the performance more precisely, we used object popularity weighted precision (WP), recall (WR) and F$_1$ score (WF1) as additional accuracy metrics.

\begin{itemize}
\item \emph{Weighted precision}: Weighted by the popularity of objects, WP is calculated as:
\begin{equation}
\label{equa:21}
weighted \, precision=\frac{1}{K}\sum_{k=1}^{K}{\sum_{n=1}^{|\mathcal{O}|}{\frac{|{\mathcal{V}_o}^{*(k)} \cap {\mathcal{V}_o}^g|}{|{\mathcal{V}_o}^{*(k)}|}}}\cdot \mathcal{I}_o
\end{equation}
\item \emph{Weighed recall}: similarly, WR is calculated as:
\begin{equation}
\label{equa:22}
weighted \, recall=\frac{1}{K}\sum_{k=1}^{K}{\sum_{n=1}^{|\mathcal{O}|}{\frac{|{\mathcal{V}_o}^{*(k)} \cap {\mathcal{V}_o}^g|}{|{\mathcal{V}_o}^g|}}}\cdot \mathcal{I}_o
\end{equation}
\item \emph{Weighted F$_1$ score}: it is the harmonious mean of weighted precision and weighted recall, also computed by applying Equation~\eqref{equa:19}.
\end{itemize}

\subsection{Performance Comparison}
\label{subsec:Comparative Studies}
Table~\ref{tab:methods_Comparison} shows the performance of different methods on the two real-world datasets in terms of accuracy and efficiency. For all the accuracy evaluation metrics except precision, SmartMTD consistently achieved the best results. Even in terms of precision, SmartMTD still showed the second best performance for the experimental datasets. Among the four methods specially designed for MTD, our approach is the most efficient as demonstrated by its lowest execution time. This is because LTM and MTD-hrd include complicated Bayesian inference over the probabilistic graphical model 
and 
MBM conducts time-consuming copy detection, while our approach is based on a relatively simple graph model.

\begin{figure*}
\centering
\subfloat[][\scriptsize{Performance comparison between SmartMTD-Core and SmartMTD-C during the iteration.}]{\includegraphics[trim=0 0 0 0, clip=true, height=1.5in]{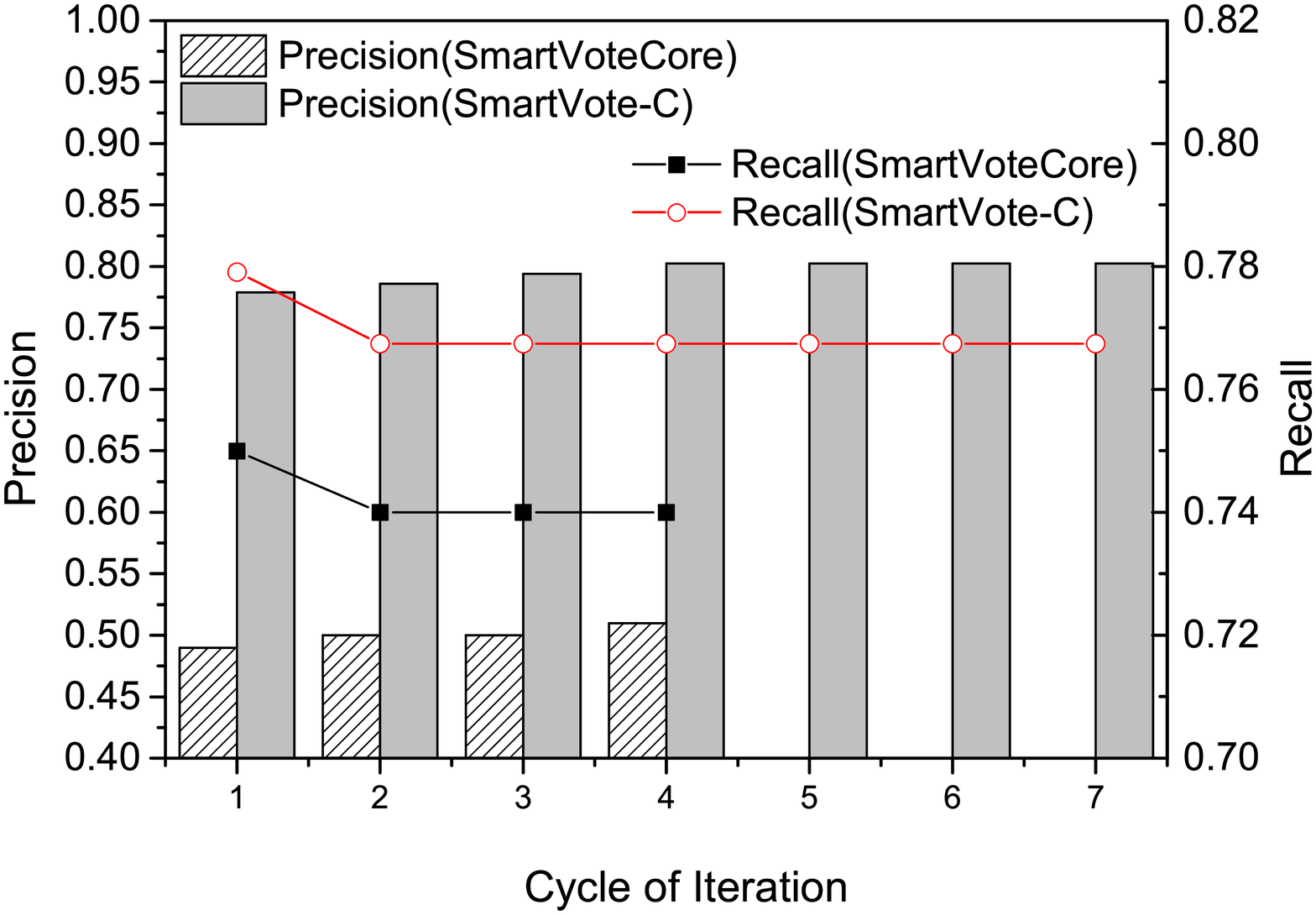}\label{fig:impact-copy}}
\hfill
\subfloat[][\scriptsize{Distribution of object popularity in Book-Author dataset}]{\includegraphics[trim=0 0 0 0, clip=true, height=1.5in]{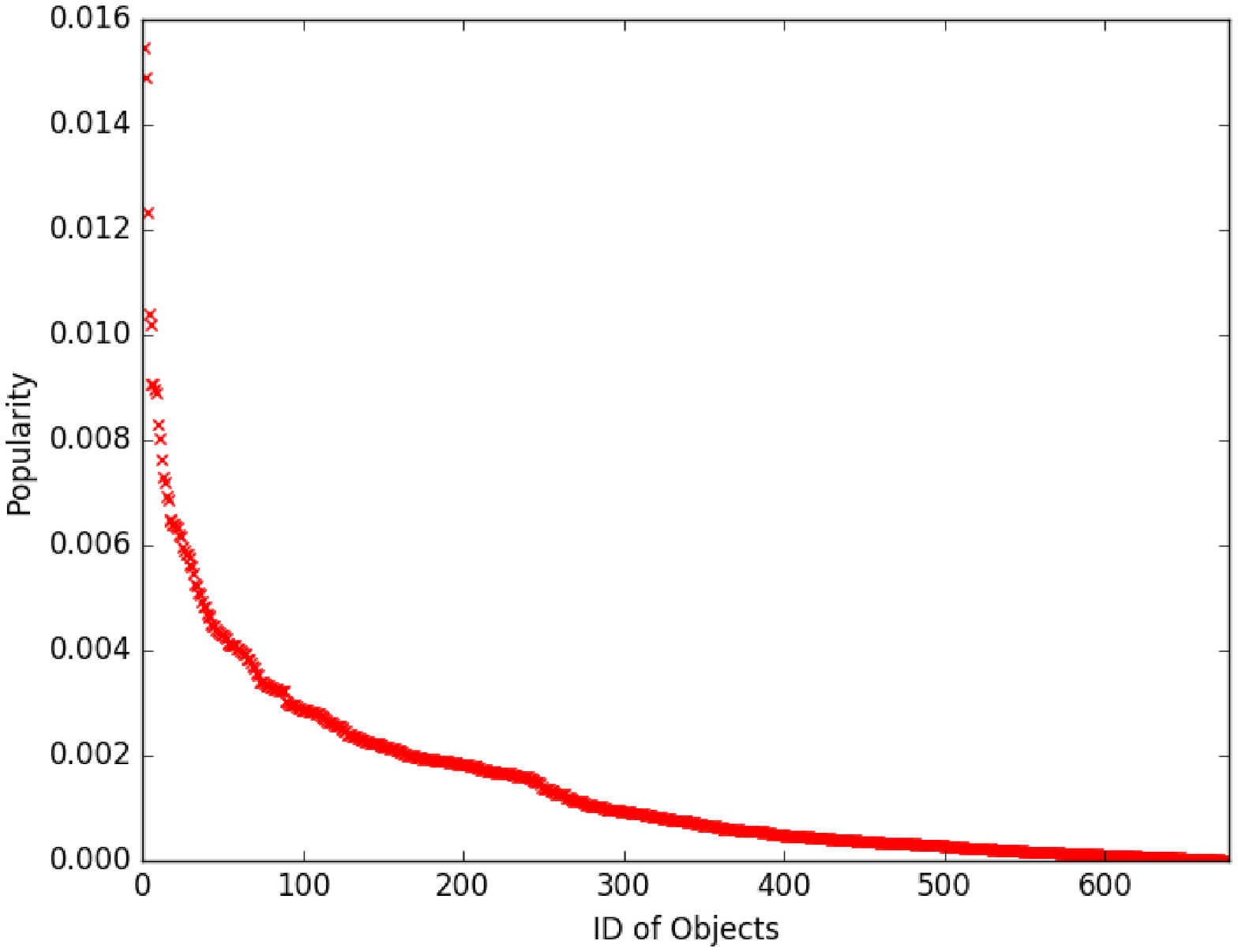}\label{fig:fig:impact-popularity-a}}
\hfill
\subfloat[][\scriptsize{Distribution of object popularity in Parent-Children dataset}]{\includegraphics[trim=0 0 0 0, clip=true, height=1.5in]{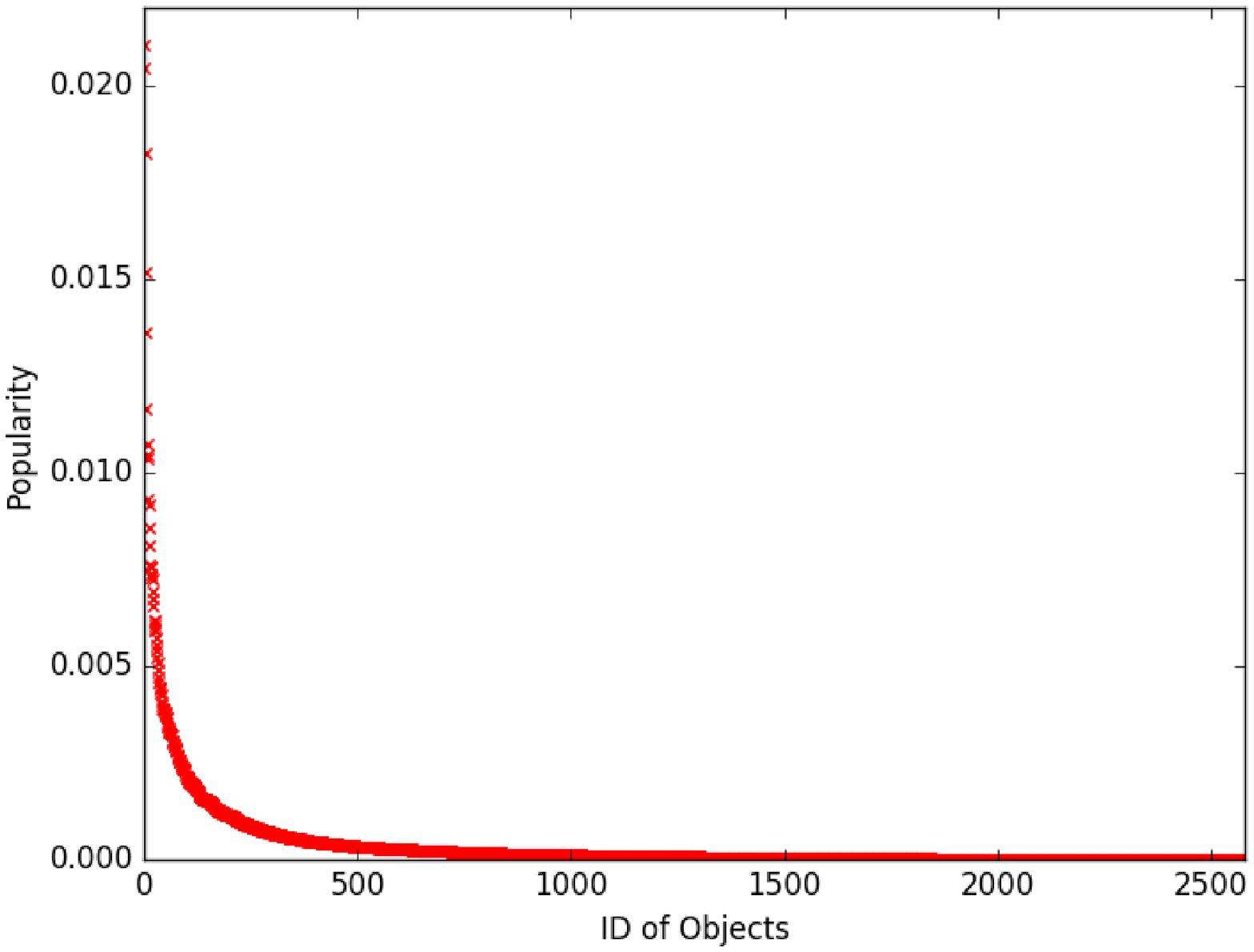}\label{fig:fig:impact-popularity-b}}
 \vspace{-2mm} 
\caption{Impact of different concerns}
 \vspace{-2mm} 
\label{fig:concerns}
\end{figure*}

All methods performed better on the Parent-Children dataset than on the Book-Author dataset. This is because the former is much bigger than the latter and provides more data for methods to make better predictions of the truth. The majority of methods showed higher precision than recall, reflecting the relatively high positive precision than the negative precision of most real-world sources. Since Voting conducts truth discovery without iteration and consideration of the quality of sources, it has the lowest accuracy but consumes the minimum execution time. Besides our approach, 2-Estimates and MBM also performed better than other methods. This can be attributed to their consideration of mutual exclusion. Though LTM and MTD-hrd also take this implication into consideration, they make strong assumptions on the prior distributions of latent variables. For this reason, once the dataset does not comply with the assumed distributions, it performs poorly. Without incorporating object popularity, 2-Estimates and MBM showed lower quality in terms of weighted measures than traditional measures. Compared with MBM, which showed the second best performance, SmartMTD not only includes object popularity but also models two types of source relations globally. Thus, SmartMTD showed the best accuracy performance.

\subsection{Impact of Different Concerns}
\label{subsec:Impact of Different Concerns}
To evaluate the impact of different implications, we implemented three variants of SmartMTD:
\begin{itemize}
\item {\em SmartMTD-Core}: A variant of SmartMTD without incorporating malicious agreement detection and object popularity quantification.

\item {\em SmartMTD-C}: A version of SmartMTD that only adopts the malicious agreement detection.

\item {\em SmartMTD-P}: A version of SmartMTD that only incorporates the object popularity quantification.
\end{itemize}

Figure~\ref{fig:impact} reports the performance comparison of SmartMTD-C, SmartMTD-P, SmartMTD, and SmartMTD-Core on the Book-Author dataset. By incorporating each implication, our approach achieved better performance 
in accuracy while only increased execution time slightly. The full version of SmartMTD, which consists of both implications, led to the best result. The experimental results on the Parent-Children dataset show the similar insights.

\vspace{2mm}
\noindent{\bf Malicious Agreement}. By detecting malicious agreement, the performance of our algorithm improved dramatically 
on precision, recall, and F$_1$ score. Interestingly, when we leveraged the weighted metrics to evaluate SmartMTD-C, the algorithm even showed better results than using traditional measures. The results reveal the wide existence of copying relations in real-world datasets. Neglecting these relations would lead to the result of over-estimating the reliability of copiers and impair the performance of truth discovery methods. Compared with object popularity quantification, malicious agreement detection is more time-consuming, as we need to compute the dependence score of each source on each object iteratively, as well as calculate the reliability of each source iteratively from the dependence scores of sources and the confidence scores of values. However, when compared with the performance improvement introduced by incorporating this implication, this additional 
time can be justified.
To further study the effect of this implication, we compared the performance of SmartMTD-Core and SmartMTD-C in terms of precision and recall for each cycle of iteration, as shown in Figure~\ref{fig:impact-copy}. The results show that although SmartMTD-C took a long time to converge, i.e., $7$ rounds of iteration (while SmartMTD only required $4$ rounds of iteration), it consistently achieved better performance in each round of iteration. 

\vspace{2mm}
\noindent{\bf Object Popularity}. By 
differing popularity of objects, our algorithm performed better in terms of accuracy with nearly no extra time. This is because more sources provide claims on popular objects, and more evidence can be obtained to model the endorsement among sources. Therefore, when computing source reliability, assigning more weights to the popular objects would lead to better truth discovery. In addition, object popularity is calculated directly from the multi-source data. Since this calculation is outside of the iteration, it can be conducted effectively under linear time. Another observation was that SmartMTD-P achieved higher weighted accuracy than traditional accuracy. This is consistent with our expectation that source reliability evaluation relies on the claims provided on popular objects. By differentiating the popularity of objects, our approach obtained results of higher precision. By ranking objects in the Book-Author dataset and the Parent-Children dataset, respectively, in a descending order of their popularity degrees, we draw scatter diagrams as shown in Figure~\ref{fig:fig:impact-popularity-a} and~\ref{fig:fig:impact-popularity-b}, where each point depicts an object with the corresponding popularity degree (totally, there are $677$ objects in the Book-Author dataset and $2,579$ objects in the Parent-Children dataset). We observed that in both 
diagrams, the points with very high popularity degrees are quite sparse, indicating that only very few objects are more popular than the majority.

To further validate SmartMTD, we compared SmartMTD with MBM (the best baseline method) on top-$20$ popular objects in the ground truth of the Book-Author dataset. SmartMTD returned false values on $2$ objects (Book $id:9780072499544$ and $id:9780071362856$) while MBM made mistakes on $4$ objects (Book $id: 9780028056005$, $id:9780072499544$, $id:9780071362856$, and $id: 9780072843996$), demonstrating that SmartMTD had better accuracy on the more popular objects. SmartMTD and MBM both returned false values on Book $id:9780072499544$ and $id:9780071362856$  because some authors are neglected by all the sources.

\vspace{-2mm}
\section{Related Work}
\label{sec:Related_Work}
Significant research efforts have been contributed to truth discovery in various application scenarios (see~\cite{li2012truth,waguih2014truth,li2015survey} for surveys). The \emph{primitive} methods are typically \emph{rule-based}, such as the methods that take the \emph{majority voting} (for categorical data) or the \emph{mean} (for continuous data) as the true values. These methods do not distinguish the reliability of sources and therefore have low accuracy when many sources provide low-quality data. 

Yin et al.~\cite{yin2008truth} first formulate the truth discovery problem in 2008. Since then, many advanced solutions have been proposed by additionally considering various implications of multi-source data. They generally fall into five categories. 
The \emph{link} based methods~\cite{pasternack2010knowing,kleinberg1999authoritative} conduct random walks on the bipartite graph between sources and values of objects. They measure source authority based on the links to the claimed values and estimate source reliability and value correctness based on the bipartite graph. \emph{Iterative} methods~\cite{yin2008truth,pasternack2010knowing,galland2010corroborating} iteratively calculate value veracity and source reliability from each other until certain convergence condition is met. \emph{Bayesian point estimation} methods~\cite{dong2009integrating,wang2015integrated} adopt \emph{Bayesian analysis} to compute the maximum a posteriori probability or \emph{MAP} value for each object. \emph{Probabilistic graphical model} based methods~\cite{zhao2012probabilistic,zhao2012bayesian,wang2016implications} apply probabilistic graphical models to jointly reason about source trustworthiness and value correctness. 
Finally, \emph{optimization} based methods~\cite{li2014resolving,li2014confidence} formulate the truth discovery problem as an optimization problem. Recently, Popat et. al~\cite{Popat2017Where} propose an approach for early detection of emerging claims, which copes with textual claims. This is a very interesting direction for truth discovery, but out of the scope of our paper.

Despite active research in the field, multi-truth discovery (MTD) is rarely studied by the previous work. LTM (Latent Truth Model)~\cite{zhao2012bayesian}, a probabilistic graphical model based method, is the first solution to the MTD problem. In this work, Zhao et al. measure two types of errors (false positive and false negative) by modeling two different aspects of source reliability (\emph{specificity} and \emph{sensitivity}) in a generative process. The disadvantage is, LTM makes strong assumptions about prior distributions for nine latent variables, rendering the model inhibitive and intractable to incorporating various implications to improve its performance. Pochampally et al.~\cite{pochampally2014fusing} study various correlations among sources by taking information extractors into consideration, the application scenario is different from ours. The experiments show that their basic model without considering source correlations sometimes performs worse than LTM, while in our experiments, SmartMTD constantly achieves considerably better results than LTM. To rebalance the distributions of positive claims and negative claims and to incorporate the implication of values' co-occurrence in the same claims, Wang et al.~\cite{wang2016implications} propose a probabilistic model that takes multi-valued objects into consideration. However, this method also requires initialization of multiple parameters, such as prior true or false count of each object, and prior false positive or true negative count of each source. Waguih et al.~\cite{waguih2014truth} conclude with extensive experiments that these probabilistic graphical model-based methods cannot scale well. Zhi et al.~\cite{zhi2015modeling} also consider the mutual exclusion between sources' positive claims and negative claims, but they model the silence rate of sources to tackle the possible non-truth objects rather than multi-valued objects. To relax unnecessary assumptions, Wang et al.~\cite{wang2015integrated} analyze the unique features of MTD and propose an MBM (Multi-truth Bayesian Model), which incorporates source confidence and finer-grained copy detection techniques in a Bayesian framework. However, they assume that false information is copied among sources and correct information is provided independently by sources. Recently, Wang et al.~\cite{wang2016empowering} design three models (i.e., the \emph{byproduct} model, the \emph{joint} model and the \emph{synthesis} model) for enhancing existing truth discovery methods. Their experiments show that those models are effective in improving the accuracy of multi-truth discovery using existing truth discovery methods. However, LTM and MBM still performed better than those enhanced methods. Wan et al.~\cite{wan2016uncertainty} propose an uncertainty-aware approach for the real-world cases where the number of true values is unknown. However, they cope with continuous data rather than categorical data.

Distinguishing from the above MTD methods, our 
SmartMTD features a graph-based approach,
which has three novel features: i) while \emph{object difficulty}~\cite{galland2010corroborating} (i.e., the difficulty of getting true values for each object) and \emph{object relations}~\cite{pasternack2010knowing,Yu2014minority} (i.e., objects may affect each other) have been studied by the previous work, SmartMTD creatively considers the impact of object popularity on source reliability; ii) instead of assuming independence of sources (in LTM) or independent copying relations among sources (like in MBM), SmartMTD models copying relations globally by constructing graphs of all sources that provide values on a specific object; iii) in addition to the copying relations among sources, which are the only source relations considered by MBM and other methods~\cite{dong2009integrating}, SmartMTD not only punishes the \emph{malicious copiers} that make the same faults as the sources from which they copy but also defines a new source relation, named \emph{supportive relation} to describe sources' implicit support for each other in providing the same true values.

\section{Conclusions}
\label{sec:Conclusion}
In this paper, we focus on the problem of truth discovery for multi-valued objects (or MTD), which has rarely been studied by the previous efforts. We propose a graph-based approach, called \emph{SmartMTD}, which incorporates two important concepts, namely \emph{source relations} (including \emph{supportive relations} and \emph{copying relations}) and \emph{object popularity}, for better truth discovery. In particular, we construct $\pm$\emph{supportive agreement graph}s to model the endorsement among sources on their positive and negative claims, from which two aspects of source reliability (i.e., positive precision and negative precision) are derived. Copying relations among sources are captured by constructing the $\pm$\emph{malicious agreement graph}s based on the consideration that sources sharing the same false values are more likely to be dependent. We also consider the impact of the popularity of objects on source reliability calculation. We develop techniques to quantify object popularity based on object occurrences and source coverage. Experimental results on two large real-world datasets show that our approach outperforms the state-of-the-art truth discovery methods.

Our future work will focus on improving the proposed graph-based model by exploring more implications such as the long-tail phenomenon on source coverage and differed source confidence on positive and negative claims. We will also conduct more experimental studies to further validate the performance of SmartMTD.